\documentclass[12pt]{iopart}
\begin{document}
\title{Scale invariant gravity and the quasi-static universe}
\author{Robin Booth\\ \small{Theoretical Physics, The Blackett Laboratory}
\\ \small{Imperial College, Prince Consort Road, London SW7 2BZ, UK}}


\maketitle

\begin{abstract}
We highlight the fact that the lack of scale invariance in the
gravitational field equations of General Relativity results from
the underlying assumption that the appropriate scale for the
gravitational force should be linked to the atomic scale. We show
that many of the problems associated with cosmology and quantum
gravity follow directly from this assumption.  An alternative
scale invariant paradigm is proposed, in which the appropriate
scale for General Relativity takes the Universe as its baseline,
and the gravitational force does not have any fixed relationship
to forces that apply on the atomic scale. It is shown that this
gives rise to a quasi-static universe, and that the predicted
behaviour of this model can resolve most of the problems
associated with the standard Big Bang model. The replacement of
Newton's gravitational constant in the quasi-static model by a
scale-dependent renormalisation factor is also able to account for
a number of astronomical observations that would otherwise require
ad-hoc explanations. Some of the implications of scale invariant
gravity for Planck scale physics, quantum cosmology, and the
nature of time are discussed.
\end{abstract}

\section{Introduction}
Einstein's General Theory of Relativity has proved to be one of
the most successful and enduring theories in physics, and its
predictions have been verified in numerous experiments.  However,
it stands alone amongst field theories in that it is not scale
invariant.  For example, the differential form of Maxwell's
equations, which elegantly describe the electromagnetic field, do
not define any intrinsic scale.  Conversely, Einstein's field
equations, which describe the way that matter curves spacetime,
are linked to an apparently arbitrary scale determined by the
Newtonian gravitational constant, $G$.

This can be verified by inspection of the standard
Einstein-Hilbert action for General Relativity

\begin{equation}\label{eqn:GRaction}
    S = \frac{1}{{16\pi G}}\int {d^4 x\sqrt { - g} R
    + \int {d^4 x\sqrt { - g} } L_M \left( \phi  \right)}
\end{equation}

In the case of a universe without matter, only the curvature term
in the action will be applicable.  This can readily be made scale
invariant by removing the $16 \pi G$ factor using the
transformation $g'_{\alpha\beta}\rightarrow \Omega^2
g_{\alpha\beta}$, $R' \rightarrow \Omega^{-2}R$, where $\Omega$ is
a constant. Similarly, if the matter Lagrangian term is considered
in isolation, it can be shown that any constituent fields that
obey the Dirac equation, for example the electromagnetic field,
will be also be invariant with respect to a conformal
transformation. However, when the gravitational and matter terms
are combined as in (\ref{eqn:GRaction}), the resulting action is
no longer scale invariant.

The scale dependency inherent in General Relativity leads to an
apparently fundamental scale, defined by the Planck units:

\paragraph{Planck Length}
\begin{equation}\label{Lplanck}
{\rm  }L_P  \equiv \sqrt {\frac{{\hbar G}}{{c^3 }}} \simeq
10^{-35}m
\end{equation}

\paragraph{Planck Time}
\begin{equation}\label{Tplanck}
 T_P  \equiv  \frac{{L_P }}{c} \simeq 10^{-43}s
\end{equation}

\paragraph{Planck Mass}

\begin{equation}\label{Mplanck}
M_P  \equiv  \sqrt {\frac{{\hbar c}}{G}} \simeq 10^{-8}kg
\end{equation}

No theory of Quantum Gravity is yet able to explain the wide
disparity between the Planck scale and the atomic scales that
govern the everyday world we inhabit.  Whilst the magnitude of the
Planck length and time scales, at $\simeq 10^{-20}$ of the
corresponding atomic length and time scales, are just about
reconcilable with the concept of an evolving quantum Universe, it
is difficult to account for the fact that the Planck mass is
$\simeq 10^{20}$ times larger than the proton mass.

\par The remainder of this introductory section provides a brief
historical overview of the development of alternative formulations
of gravity.  In Section \ref{sec:Plankproblems} we review some of
the main problems that exist in cosmology and quantum gravity, and
show that these follow as a direct consequence of the lack of
scale invariance in General Relativity. Section
\ref{sec:scalefree} introduces an alternative approach to
scale-invariant gravity, based on the redefinition of the
gravitational field equations using cosmological units. It is
shown that the resulting dynamical equations describe a universe
that can be either static or expanding, depending on the reference
frame of the observer. In Section \ref{sec:consequences}, some of
the consequences of the scale invariant version of General
Relativity described in Section \ref{sec:scalefree} are examined.
It is shown that the modified dynamical behaviour can
satisfactorily account for the problems associated with Big Bang
cosmology. Section \ref{sec:conclusions} concludes by examining
some of the implications of abandoning the Planck scale for the
study of quantum gravity.

Numerous attempts have been made to develop a theory of
gravitation that is scale invariant, and yet retains the key
properties of General Relativity, such as the principle of general
covariance.  These can be classified according to their main
approach to the problem:
\begin{itemize}
  \item Adding a scalar field to the standard Einstein-Hilbert
  action
  \item Resorting to a higher order description of gravity in place of the
  Riemann tensor of General Relativity
  \item Modifying the behaviour of the fields that constitute the
  matter Lagrangian
\end{itemize}
There are also a number of hybrid theories that incorporate more
than one of the above approaches.

\paragraph{Scalar-Tensor theories} Probably the best known of this class
of theories is the Brans-Dicke theory \cite{dicke:mach}.  The
motivation for the development of this variant of General
Relativity was to create a theory of gravity that explicitly
incorporated Mach's principle, and the Dirac Large Number
Hypothesis (LNH) \cite{dirac:1938}, whilst still retaining the
essential symmetric and divergence-free nature of the original
Einstein field equations. Although the Brans-Dicke theory does not
directly address the issue of scale invariance, it is of
particular interest in the historical development of this subject
as it derives from the underlying Mach's principle: that inertial
mass results from the gravitational interaction between matter and
all other matter in the universe, and that the gravitational
constant is itself determined by the matter distribution in the
Universe, so that ${{GM} \mathord{\left/ {\vphantom {{GM} {Rc^2
\sim 1}}} \right. \kern-\nulldelimiterspace} {Rc^2  \sim 1}} $.
Accordingly, either $G$ or $M$ must vary over time as $R$
increases. In the introduction to \cite{dicke:mach}, the authors
remark on the fact that the metric tensor is dependent on the
choice of units used in any particular representation of the
gravitational field equations. The units chosen in General
Relativity are such that nucleons have physical properties that
are independent of location (and time).
\par In the Brans-Dicke theory the gravitational constant in the Einstein field equation
is replaced by a scalar field \( \phi \sim {1 \mathord{\left/
 {\vphantom {1 G}} \right.
 \kern-\nulldelimiterspace} G}
\).  In order to preserve the divergence-free nature of the LHS of
the field equation, such that \( {T^{\mu \nu }} _{;\mu }  = 0 \),
it is necessary to introduce various derivatives of the scalar
field, together with a coupling constant $\omega$, to give the
Brans-Dicke action

\begin{equation}\label{eqn:BDaction}
S = \int {d^4 x\sqrt { - g} R\left[ {\frac{{\phi R}}{{16\pi
}}{\cal {L}}_m - \frac{\omega }{\phi }\nabla _\mu  \phi \nabla
^\mu \phi } \right]}
\end{equation}

Varying (\ref{eqn:BDaction}) with respect to the metric gives the
modified field equation for Brans-Dicke gravity

\begin{equation}\label{eqn:BDgravity}
G_{\mu \nu }  = \frac{{8\pi }}{\phi }T_{\mu \nu }  - \frac{{\omega
^2 }}{\phi }\left( {\nabla _\mu  \phi \nabla _\nu \phi  - g_{\mu
\nu } \nabla _\alpha  \phi \nabla ^\alpha  \phi } \right) -
\frac{1}{\phi }\left( {\nabla _\mu  \phi \nabla _\nu \phi  -
g_{\mu \nu }  \partial ^\alpha  \partial _\alpha  \phi } \right)
\end{equation}

It is worth noting that the scalar field $\phi$ in the Brans-Dicke
theory is analogous to the dilaton field  $\phi$ that arises in
formulations of gravity derived from string theory.

\par Other applications of the scalar-tensor approach include the Conformal General
Relativity model proposed in \cite{pervushin:2001}, and the
Non-Gravitating Vacuum Energy model (NGVE) \cite{guendelman:2001}.
The former incorporates a conformal invariant
$\hat{g}_{\mu\nu}=W^2g_{\mu\nu}$, where $W$ is a dilaton scalar
field described by the Penrose-Chernikov-Tagirov (PCT) action

\begin{equation}
  I=-\int d^4x (-\hat{g})^{1/2}R(\hat{g})/6
\end{equation}

The NGVE model features a scalar density in addition to the
standard Lagrangian of General Relativity.

\paragraph{Higher order gravity theories} The standard theory of General
Relativity is based on the second order Einstein-Hilbert action of
(\ref{eqn:GRaction}), which is formed from the Riemann curvature
tensor.  An alternative form of gravity was proposed by Weyl in
1918, in which the Einstein-Hilbert action is replaced with a
conformal invariant fourth order action

\begin{equation}\label{eqn:Weyl_action}
I_W = -\alpha \int d^4x (-g)^{1/2}
C_{\lambda\mu\nu\kappa}C^{\lambda\mu \nu\kappa}
\end{equation}

where $C_{\lambda\mu\nu\kappa}$ is the conformal Weyl tensor and
$\alpha$ is a purely dimensionless coefficient.  This formulation
was used as the basis for the Conformal Cosmology model developed
by Mannheim \cite{mannheim:1989,mannheim:1990}. Conformal gravity
possesses no intrinsic fundamental length scale. The gravitational
equations must be fourth order, so standard the Einstein equation
is replaced by
\begin{equation}
    4 \alpha W_{\mu \nu} =4\alpha (W^{(2)}_{\mu\nu} - W^{(1)}_{\mu\nu}/3)=T_{\mu\nu}
\end{equation}

where $W_{\mu  \nu}$ is given by
\begin{eqnarray}
    W^{(1)}_{\mu\nu} & = & 2g_{\mu\nu}(R^{\alpha}_{\phantom{\alpha}\alpha})
    ^{;\beta}_{\phantom{;\beta};\beta} -
    2(R^{\alpha}_{\phantom{\alpha}\alpha})_{;\mu;\nu} -2
    R^{\alpha}_{\phantom{\alpha}\alpha} R_{\mu\nu}
    +g_{\mu\nu}(R^{\alpha}_{\phantom{\alpha}\alpha})^2/2 \\
    W^{(2)}_{\mu\nu} & = & g_{\mu\nu}(R^{\alpha}_
    {\phantom{\alpha}\alpha}) ^{;\beta} _{\phantom{;\beta};\beta}/2  +
    R_{\mu\nu\phantom{;\beta};\beta}^{\phantom{\mu\nu};\beta}
    - R_{\mu\phantom{\beta};\nu;\beta}^{\phantom{\mu}\beta}
    -R_{\nu\phantom{\beta};\mu;\beta}^{\phantom{\nu}\beta}
    - 2R_{\mu\beta}R_{\nu}^{\phantom{\nu}\beta}
    +g_{\mu\nu}R_{\alpha\beta}R^{\alpha\beta}/2
\end{eqnarray}

\paragraph{Modified matter Lagrangian theories} The most notable
theory of this type is the steady-state model of Hoyle and
Narlikar \cite{hoyle}. Essentially, this involves particle masses
increasing as the Universe expands, and the existence of a scalar
field (the C-field) that is capable of continuously creating new
particles. The creation of new matter exactly balances the
decrease in matter density resulting from the expansion of the
Universe in such a way that the energy density in the Universe
remains constant, and the gravitational equations become scale
invariant.

\paragraph{Hybrid theories} An example of a theory that combines
two of the above approaches is the Scale-Covariant Theory of
Gravitation proposed by Canuto et al
\cite{canuto:1977,canuto:1978}.  This incorporates an additional
tensor term, together with a time dependent scalar factor, in the
gravitational action.  The action is then fourth order, and has
similar scale invariant properties to the Weyl action.

As a generalisation, it can be said that all three of these
approaches have the ability to address specific problems
associated with the lack of scale invariance in General
Relativity, but only at the expense of abandoning the elegance of
Einstein's original theory. Furthermore, none of these approaches
is able to provide a comprehensive framework that resolves all the
issues identified in the following section.

\section{The Planck scale crisis}\label{sec:Plankproblems}
The objective of this section is to illustrate how the lack of
scale invariance in the field equations of General Relativity
leads ultimately to a range of cosmological problems, and a crisis
at the Planck scale.  The consequences of this scale invariance
can be broadly broken down into four area:
\begin{itemize}
  \item Cosmological dynamics
  \item Singularities
  \item Energy conservation
  \item Time
\end{itemize}

\subsection{Cosmic dynamics}\label{sec:dynamics1}
The dynamical behaviour of the Universe is conventionally
determined by applying the Einstein field equations to the
Robinson-Walker metric in order to derive the Friedman equations.
Although this treatment is quite standard, and described in
numerous references, e.g. \cite{goobar}, the main steps are shown
here in order to highlight the point at which the assumption is
made that results in the Planck scale crisis.

\subsubsection{The Friedman equations}

Starting with the Einstein equation, derived from applying the
variation principle to the Einstein-Hilbert action of
(\ref{eqn:GRaction})
\begin{equation}\label{einstein}
G_{\mu \nu }  = \frac{{8\pi G}}{{c^4 }}T_{\mu \nu }  + \Lambda
g_{\mu \nu }
\end{equation}

and the Robinson-Walker metric
\begin{equation}\label{FLRW}
ds^2  = dt^2  - a^2 (t)\left( {\frac{{dr^2 }}{{1 - kr^2 }} + r^2
d\theta ^2  + r^2 \sin ^2 \theta d\phi ^2 } \right)
\end{equation}

we can derive the first two components of the Einstein tensor

\begin{eqnarray}\label{g00}
    G_{00} & = & \frac{{3\dot a^2 }}{{a^2 c^2 }} + \frac{{3k}}{{a^2
    }}\\
    G_{11} & = & - \frac{{k + {{2a\ddot a} \mathord{\left/
 {\vphantom {{2a\ddot a} c}} \right.
 \kern-\nulldelimiterspace} c}^2  + {{\dot a^2 } \mathord{\left/
 {\vphantom {{\dot a^2 } {c^2 }}} \right.
 \kern-\nulldelimiterspace} {c^2 }}}}{{1 - kr^2 }} \label{g11}
\end{eqnarray}

The equations of motion are derived by combining
(\ref{g00},\ref{g11}) with the corresponding 00 and 11 components
of the stress-energy tensor to give

\begin{eqnarray}\label{d2}
\frac{{3\dot a^2 }}{{a^2 }} + \frac{{3k}}{{a^2 }} = 8\pi G\rho +
c^2 \Lambda
  \\
  \frac{{2\ddot a}}{a} + \frac{{\dot a^2 }}{{a^2 }} + \frac{k}{{a^2 }}  =  - 8\pi Gp - c^2
  \Lambda \label{d3}
\end{eqnarray}

It is at this point that the assumption inherent in equation
(\ref{einstein}) manifests itself.  Implicitly, in applying the
Einstein equation, it is assumed that the energy density $\rho$ is
measured in conventional units that are defined in terms of the
atomic properties of matter, e.g. MKS units.  Similarly, it is
assumed that the scale of the spacetime curvature induced by this
stress-energy is determined by the Newtonian gravitational
constant $G$, expressed in the same units.  By expressing the
Einstein equation in this way, we are essentially saying that the
scale of a cosmological phenomenon - the curvature of spacetime -
is determined by behaviour of matter at atomic scales.  Not only
is this counter-intuitive, but as we shall see, this assumption
leads directly to the problems and paradoxes that currently beset
cosmology and quantum gravity. Restating equation (\ref{einstein})
in Planck units, by setting $G=c=1$, does nothing to overcome this
objection, assuming that we retain the direct relationship between
the Planck scale and the atomic scale.

\par Returning to the standard derivation of the Friedman equations,
since $p$ is small in the present epoch, and ignoring for now the
cosmological constant, (\ref{d3}) becomes

\begin{equation}
\frac{{2\ddot a}}{a} + \frac{{\dot a^2 }}{{a^2 }} + \frac{k}{{a^2
}} = 0
\end{equation}

and (\ref{d2}) simplifies to

\begin{equation}\label{eqn:expand}
\frac{{3\dot a^2 }}{{a^2 }} + \frac{{3k}}{{a^2 }} = 8\pi G\rho
\end{equation}

One solution to equation (\ref{eqn:expand}) for a matter dominated
universe is the Einstein-de Sitter model, with
\begin{equation}\label{Einstein-deSitter}
a(t) = a_0 \left( {\frac{t}{{t_0 }}} \right)^{\frac{2}{3}}
\end{equation}

It is conventional to define a critical energy density
\begin{equation}\label{critical}
\rho _{crit}  = \frac{{3H^2 }}{{8\pi G}}
\end{equation}

and to express the energy content of the Universe as a fraction of
this critical density
\[
\Omega  \equiv {\rho  \mathord{\left/
 {\vphantom {\rho  {\rho _c }}} \right.
 \kern-\nulldelimiterspace} {\rho _c }}
\]

\subsubsection{Age of the Universe}
 From (\ref{d2}), and ignoring for now the cosmological constant,
 $\Lambda$, an expression can be derived for the age of the universe

\begin{equation} \label{age}
  t = \int_0^{a(t)} {\frac{{da}}{{\sqrt {{{8\pi G\rho _0 a^3 \left( {t_0 } \right)} \mathord{\left/
 {\vphantom {{8\pi G\rho _0 a^3 \left( {t_0 } \right)} {3a\left( {t} \right) - kc^2 }}} \right.
 \kern-\nulldelimiterspace} {3a\left( {t} \right) - kc^2 }}} }}}
\end{equation}

which, in the case of a flat universe with $k=0$, evaluates to
\begin{equation} \label{age2}
t_0  = \frac{2}{3}H_0
\end{equation}

Current measurements of the Hubble constant from low-redshift Type
Ia supernovae \cite{hamuy:1996} give a value of $H_0 = 63.1 \pm
4.5 km s^{-1} Mpc^{-1}$.  Based on this same data, the standard
Big Bang theory would (with $\Lambda =0$) give a value for the age
of the universe of $t_0 \approx 10 \times 10^9$ years, which is at
odds with astrophysical and geological evidence.

\subsubsection{Big Bang problems}\label{sec:bigbang}
The standard cosmological problems are well documented in a number
of sources, including \cite{jm:1}.  We show here how they arise
directly from the assumptions made in the preceding section, and
also highlight an additional problem which has not previously been
noted in the literature on this subject.

\paragraph{The horizon problem}
The standard theory is unable to explain how regions of the
Universe that had not been in contact with each other since the
Big Bang are observed to emit cosmic background radiation at
almost precisely the same temperature as each other.

\paragraph{The flatness problem}
The fact that the observed matter density of the Universe is so
close to the critical value necessary for the Universe to be
closed implies that the ratio of these densities, $\Omega$, must
have been very close to one at the time of the Big Bang. This
observation is the one of the main reasons for the widely held
belief that there is probably additional hidden mass in the
Universe such that $\Omega$ is in fact equal to one in the present
epoch.  The standard theory is unable to provide an explanation as
to why the density of the Universe should be so close to the
critical value.

\paragraph{The lambda problem}
Einstein originally included the cosmological constant $\Lambda$
in his gravitational field equations in order to arrive at a
solution that was consistent with the prevalent concept of a
static Universe. The subsequent discovery that the Universe is in
fact expanding has done nothing to diminish the enthusiasm on the
part of many theorists for retaining this constant, in spite of
Einstein's opinion that its inclusion was his `biggest blunder'.
Recent measurements of the expansion rate of the Universe
\cite{reiss:supernovae} appear to suggest that a non-zero
$\Lambda$ may be causing the expansion to accelerate.  However,
the expansion of the Universe would have caused any initial
cosmological constant to grow by a factor of $10^{128}$ since the
Plank epoch. For $\Lambda$ to be as small as it appears today
presents yet another fine tuning problem.

\paragraph{The linearity problem}
For higher redshifts, the Hubble diagram can tell us whether the
expansion of the Universe has undergone any periods of
acceleration in its history. These results are derived using the
expression for the expansion rate in terms of the redshift
\begin{equation}
 H^2  = H_0^2 \left[ {\Omega _M \left( {1 + z} \right)^3  + \Omega _K \left( {1 + z} \right)^2  + \Omega _\Lambda  } \right] \\
\end{equation}

where

\[
\begin{array}{l}
 \Omega _M  \equiv \left( {\frac{{8\pi G}}{{3H_0^2 }}} \right)\rho _0  \\
 \Omega _\Lambda   \equiv \frac{\Lambda }{{3H_0^2 }} \\
 \Omega _K  \equiv \frac{{ - k}}{{a_0^2 H_0^2 }} \\
 \end{array}
\]
and
\[
\Omega _M  + \Omega _\Lambda   + \Omega _K  = 1
\]

This leads to a formula for the lookback time in terms of the
present value of the Hubble parameter, and the red-shift (see
\cite{goobar} for  derivation)
\begin{equation}\label{lookback}
  t_0  - t_1  = H_0^{ - 1} \int_0^{z{}_1} {(1 + z)^{ - 1} \left[ {\left( {1 + z} \right)^2 \left( {1 + \Omega _M z} \right) - z(2 + z)\Omega _\Lambda  } \right]^{ - {\textstyle{1 \over 2}}} dz}  \\
\end{equation}

Analysis of the data obtained from high redshift supernovae
\cite{reiss:supernovae} gives best fit values of $\Omega_M=0.28,
\Omega_\Lambda=0.72$, and an age for the Universe of $H_0^{ -
1}\simeq 15 \times 10^9 $ years.  The values of the density
parameters are remarkable in that they are very close to the
values that would give a best fit curve corresponding to a
linearly expanding universe of the same age, characterised by
$t_0=1/H_0$.  The fact that this is the case may merely be one of
the many coincidences that appear to exist in cosmology.
Alternatively, it could be an indication that some underlying
mechanism is in place that not only ensures that $\Omega_{Tot}=1$,
but also adjusts the cosmological constant over time such that
$t_0=1/H_0$ will always apply.

\subsection{Singularities}
The standard formulation of the Einstein field equations leads
inevitably to the concept of singularities: regions of space where
the mutual gravitational energy of a body of matter becomes
infinitely large, but is contained within an infinitesimally small
volume.  Standard Big Bang cosmology gives rise to singularities
in at least two sets of circumstances: the initial Big Bang
itself, and the interior of Black Holes.  The origin of the Big
Bang singularity follows directly from the dynamical equations
described in Section \ref{sec:dynamics1}.  If the expansion of the
Universe were to be reversed, then it ie easy to see that there
must have been a time when all the matter/energy in the Unverse
was contained in a vanishingly small volume of space.  The proof
that all Black Holes must contain a singularity is somewhat less
intuitive, and was first achieved using topological methods in the
Penrose-Hawking singularity theorem \cite{hawking:1970}.  The
concept of a singularity is abhorrent to most physicists,
involving as it does the existence of infinities.  Typically, the
worst implications of singularities are avoided by invoking the
Planck scale, with the assumption that the standard laws of
physics somehow break down at energies greater than the Planck
energy, or at distances less than the Planck length. However, this
get-out is intellectually far from satisfactory.  There is no
obvious reason why a different set of physical laws should be
applicable at the Planck scale. Nevertheless, the presumption that
this is the case provides one of the main motivations for the
study of Quantum Gravity.

\subsection{Energy conservation}
Mach's Principle, in its most basic form, asserts that the inertia
experienced by a body results from the combined gravitational
effects of all the matter in the Universe acting on it.  Although
a great admirer of Mach, Einstein was never entirely certain
whether General Relativity incorporated Mach's Principle.  Indeed,
the issue is still the subject of continued debate even today (see
\cite{barbour:mach} for example). A stronger version of Mach's
Principle can be formulated, which states that the inertial mass
energy of a matter particle is equal and opposite to the sum of
the gravitational potential energy between the particle and all
other matter in the Universe, such that:
\begin{eqnarray}
    mc^2  & = & - \sum\limits_N {\frac{{Gm.m}}{r}}  \\
    &= &-4\pi \alpha Gm\int\limits_0^R {\rho (r)r^2
    .\frac{1}{r}dr}\label{eqn:mach}
\end{eqnarray}

where $R \equiv c/H$ is the gravitational radius of the Universe,
and $\alpha$ is a dimensionless constant.  In the case of a
homogeneous and isotropic matter distribution this becomes

\begin{equation}
    mc^2  = -2\pi \alpha Gm \bar \rho R^2 \label{eqn:mach2}
\end{equation}

where $ \bar \rho $ is the average matter density of the Universe.

\par Observational evidence suggests that the relationship  \( {{G\rho
_0 } \mathord{\left/
 {\vphantom {{G\rho _0 } {H_0^2  \simeq 1}}} \right.
 \kern-\nulldelimiterspace} {H_0^2  \simeq 1}}
\) is valid to a reasonable degree of precision in the present
epoch.  It would be particularly satisfying if this relationship
were to be found to be true, as it would tie in with the concept
that the Universe is `a free lunch', i.e. all the matter in the
Universe could be created out of nothing, with a zero net energy.
However, it can be seen from (\ref{eqn:mach2}) that rest mass
energy appears to be proportional to $1/R$. Consequently, as the
Universe continues to expand, the energy arising from mutual
gravitational attraction will ultimately tend towards zero.
Conversely, gravitational energy will become infinite at $t=0$,
the initial singularity. Since there is no suggestion that the
rest mass energy associated with the matter in the Universe
changes over time (unless one is considering the Steady State
Theory), it would appear that this neat zero energy condition in
the present era is just a coincidence.

\subsection{The problem of time}
The theories of Quantum Mechanics and classical General Relativity
describe physics on vastly different energy and length scales.
Nevertheless, they both have some underlying factors in common.
For example, they are both based on the assumption that spacetime
exists in the form of a four-dimensional differentiable manifold,
with a Lorentzian metric structure. However, the role of time
differs significantly between the two theories. This is not itself
a problem when dealing with physics that lies solely within either
of these domains. The problem of time really only manifests itself
when attempts are made bring these two theories together into a
theory of Quantum Gravity that can be applied to physics at the
Planck scale.  It is at this point that the difficulty reconciling
the two different descriptions of time becomes apparent.  The main
arguments are described in some detail by Isham and Butterfield in
\cite{callender}, and an analysis of some aspects of the problem
of time is given by Kucha\u{r} in \cite{butterfield}. In Quantum
Mechanics, time plays an essential role as a background parameter
in labelling the state of a quantum system; for example the time
at which a measurement is made. Conversely, time in General
Relativity is more closely integrated into the very structure of
the theory itself. Essentially, time provides the fourth dimension
of spacetime that allows a three-dimensional manifold to possess
the property of curvature. Alternatively, time can be viewed as
providing the offset between successive foliations of a
three-dimensional manifold. Interestingly, neither the General
Relativity nor the Quantum Mechanics description of time coincides
closely with our own perception, which tends to view time as a
linear flow of events with a clearly defined past, present and
future.  These concepts do not really exist at all in General
Relativity time, and only do so to a limited extent in Quantum
Mechanics.

\section{Scale invariant gravity}\label{sec:scalefree}
The Einstein-Hilbert action of General Relativity contains an
implicit assumption: that the gravitational force, which
determines the behaviour of matter on a cosmological scale, has a
fixed relationship with the fields and forces that govern the
physics of matter on atomic scales. One can see how this
assumption is compatible with a particle model of gravity, in
which the gravitational force is mediated by the exchange of
spin-2 gravitons.  However, in any theory that treats gravity as a
geometrical property of spacetime, this assumption appears to be
seriously flawed.  Indeed, it appears to be yet another example of
anthropocentric thinking, as unreasonable in its own way as the
pre-Copernican assertion that the Earth lies at the centre of the
Universe. It may therefore be instructive to look at the
consequences of breaking this linkage between the cosmological and
quantum scales.

\subsection{The Einstein equations revisited} \label{sec:newGR}
If we are to do away with the Planck scale, then this immediately
poses the problem of what to put in its place as the appropriate
scale for dealing with gravitational phenomena.  Arguably the
simplest answer is to hypothesise that, since General Relativity
describes the behaviour of the universe as a whole, the
appropriate scale is that of the Universe itself.  To make this
work, it is helpful to define a universe as being a four
dimensional manifold possessing the properties of compactness and
closure. This in turn implies that the Universe is finite in
extent.  It is convenient, but not essential, to assume that the
Universe has the topology of a 3-sphere. It is then possible to
define a length scale in which the radius of the Universe $R$ is
taken to be unity, and the Gaussian curvature is given by
$1/R^2=1$. Similarly, a universe is taken to have a unit mass, and
to contain a finite number of matter quanta, $N$.  For the sake of
simplicity in the discussion that follows, it will be assumed that
the matter content of the Universe is purely baryonic in nature.
The mass of a single quantum of matter will therefore be given by
$m=1/N$ in these new units.

\par The basic principle underlying General Relativity is retained:
that stress-energy induces curvature in the space-time manifold.
However, we now rewrite the gravitational field equations in a
form that follows directly from the definition of a universe given
above, so that

\begin{equation}\label{eqn:GR1}
G_{\mu \nu }  = T_{\mu \nu }
\end{equation}

where $T_{\mu \nu }$ is the stress-energy tensor in dimensionless
normalised units, with the total stress-energy density of the
universe as a whole being unity. If we wish to express the
stress-energy tensor and curvature radius in more conventional
units related to atomic measurement scales, then we need to apply
appropriate re-normalisation factors. The curvature must be
multiplied by $1/R^2$, where $R$ is the radius of the Universe in
our chosen length units, and the stress-energy tensor must be
divided by the average stress-energy density of the universe as a
whole, $\bar \rho c^2$. The gravitational field equation now takes
on the form

\begin{equation}\label{eqn:GR2}
G_{\mu \nu }  = \frac{{3}}{{R^2 \bar \rho c^2 }}T_{\mu \nu }
\end{equation}

It can be seen that by adopting this modified form for the
gravitational field equations, that the space-time curvature
arising from a given body of matter will depend on its scale
relative to the Universe as a whole, and not solely on its mass.
In the following section we shall see that this apparently minor
change in definition has far reaching consequences.

\par It is worth noting that it is also possible to derive the
modified gravitational field equation of (\ref{eqn:GR2}) using
Mach's Principle.  This is achieved by using the energy
conservation equation (\ref{eqn:mach2}) to derive an expression
for the gravitational constant $G$, which is then substituted in
the standard Einstein equation (\ref{einstein}).  This approach is
described in more detail in \cite{booth:2001}.

\subsection{Revised cosmic dynamics}\label{sec:dynamics2}
We look first at the implications of the modified gravitational
field equations for the evolution of the Universe.  The equations
of motion are derived by combining (\ref{g00},\ref{g11}) with the
corresponding 00 and 11 components of the modified field equation
(\ref{eqn:GR2}) to give

\begin{eqnarray}\label{d4}
  \frac{{\dot a^2 }}{{a^2 }} + \frac{{kc^2 }}{{a^2 }} =  \frac{{c^2 \rho }}{{a^2\bar
  \rho
  }}\\
 - \frac{{2\ddot a}}{a} - \frac{{\dot a^2 }}{{a^2 }} -
\frac{{kc^2 }}{{a^2 }}
  = \frac{{p}}{{\bar \rho a^2 }} \label{d5}
\end{eqnarray}

Since $p$ is small in the present epoch, (\ref{d5}) becomes

\begin{equation}
\frac{{2\ddot a}}{a} + \frac{{\dot a^2 }}{{a^2 }} + \frac{{kc^2
}}{{a^2 }} = 0
\end{equation}

and (\ref{d4}) simplifies to

\begin{equation}\label{expand2}
\dot a = c\sqrt {\frac{\rho }{{\bar \rho }} - k}
\end{equation}

From this equation it can be seen that when $\rho={\bar \rho }$,
and the curvature  $k=1$, the result is a quasi-static solution
similar to the Einstein-DeSitter model, in that it has zero net
energy and $\dot a=0$. The solution is quasi-static in that the
Universe is only flat and static in a cosmological reference
frame. In regions of matter concentration where $\rho>{\bar \rho
}$, $k$ will effectively appear to be zero and the Universe will
therefore appear to be expanding to an observer in this reference
frame, with its horizon receding at the speed of light.  In the
remainder of this paper we shall refer to this as the Quasi-Static
Universe (QSU) model.
\par This equation also embodies the negative feedback mechanism
that ensures that $\rho_{mat}$ will always be equal to
$\rho_{grav}$.  If at any point $\rho_{mat}$ should exceed
$\rho_{grav}$ then this will lead to a positive $\dot a$, which
will tend to drive $\Omega\rightarrow1$.  The converse will apply
if $\rho_{mat}$ should fall below $\rho_{grav}$.

\subsection{Newton's gravitational constant}\label{sec:Newton}
In Section \ref{sec:newGR} a revised gravitational field equation
(\ref{eqn:GR2}) was constructed, which did not contain the
Newtonian gravitational constant $G$ that appears in the standard
Einstein version of the equation.  However, in the present epoch
both these equations must reduce to the Newtonian case in the weak
field limit, and must therefore be equivalent to each other:

\begin{equation}
  \frac{{3}}{{R^2 \bar \rho c^2 }}T_{\mu \nu }  = \frac{{8\pi G}}{{c^4 }}T_{\mu \nu }
\end{equation}
giving
\begin{equation}\label{eqn:G}
  G = \frac{{3c^2 }}{{8 \pi R^2 \bar \rho }}
\end{equation}

and since $t=1/H$ in this model, we can express the gravitational
'constant' as
\begin{equation}\label{eqn:G2}
  G = \frac{{3H_0^2 }}{{8 \pi \bar \rho }}
\end{equation}
where $\bar \rho$ is the mean density of the Universe in standard
atomic units.  From this it is apparent that $G$ is only a
constant in the same sense that $H_0$ is a constant, i.e. for
observers in the present epoch, as defined by atomic time. Since
$\bar \rho \propto  t^{-3}$ and $H_0^2 \propto t^{-2}$ we can see
from (\ref{eqn:G2}) that $G \propto t$ in the atomic reference
frame.  The implications of a varying $G$ will be discussed in
Section \ref{sec:varyingG}.  It is worth emphasising that the
concept of a varying gravitational constant used here differs
entirely from the phenomenological approach adopted in much of the
existing literature on the topic of variable $G$ (see
\cite{barrow:1999} for example).  The modified gravitational field
equation described here is entirely free of any arbitrary
constants when expressed in the appropriate natural cosmological
units. The apparent gravitational constant that we observe is
simply a consequence of our chosen length and mass scales relative
to this cosmological scale.

\subsection{Photon energy conservation}
In Section \ref{sec:dynamics2} we saw that, in the reference frame
appropriate to General Relativity, the Universe can be considered
to be static. As such, the fourth dimension in General Relativity
can best be viewed as being spacelike rather than timelike.  If we
now consider what happens to a photon that is emitted  by some
atomic process at a given point in time $t_1$.  At some later time
$t_2$ an observer linked to the atomic scale (such as ourselves)
would perceive that this photon had been redshifted due to the
expansion of the universe in the time $t_2-t_1$. Based on the
assumption that the energy of the photon is directly proportional
to its frequency, as given by the Planck formula $E=h \nu$, this
observer would conclude that the photon had undergone an energy
loss given by $E=h (\nu_1-\nu_2)$. If, however, the photon is
considered from the cosmological reference frame then it does not
undergo any change, since the universe is, by definition, static
in this frame. The statement that the photon has lost energy is
therefore seen to be invalid in this frame. In the absence of any
relative motion between the cosmological frame and the atomic
frame, the photon energy in the atomic frame must therefore also
be unchanged. In other words, photons will retain the same energy
that they originally possessed at the time of their emission,
whether they are red-shifted (or blue-shifted) as a result of
Doppler shift with respect to a given observer, climbing out of a
gravitational potential, or being `stretched' by the Hubble
expansion of the Universe. The observable quantity that changes in
the reference frame of the observer is the \emph{power} of the
photon.

\par Clearly, if this prediction is valid then it would have fundamental
implications across many fields of physics.  However the effects
should be experimentally verifiable.  In the field of cosmology,
the most obvious place to look for evidence of photon energy
conservation is in the cosmic microwave background (CMB). The
standard theory is based on the assumption that as the Universe
expands, the energy density due to matter decreases as
$R^{-3}(t)$, whereas the energy density due to radiation decreases
as $R^{-4}(t)$ because of the additional energy loss due to the
red-shift.  In the QSU model this remains true when looking at the
spatial energy density. However, if one is carrying out
measurements of photon energy by integrating power measurements
over a period of time that is long in relation to the time span of
the photon wavepacket (i.e. $t>>\lambda/c$), then the energy will
be found to decrease in proportion to $R^{-3}(t)$, as for the
matter case. So, for example, a CMB photon emitted when the
Universe had a temperature of $10^{9 \circ} K$, with a present day
temperature of $2.9^\circ K$, would still retain its initial
energy given by $h\nu  = kT$, rather than  $\sim10^{-9}$ of this
value. In \cite{booth:2001} an experiment is proposed for
determining photon energy from measurements of the CMB.

\subsection{About time}\label{sec:time}
One of the defining features of the Einstein gravitational field
equation (\ref{einstein}) is the fact that both sides of the
equation are symmetric, divergence-free, second rank tensors. The
stress-energy tensor $T_{\mu\nu}$ embodies the laws of energy and
momentum conservation, such that ${T^{\beta\alpha}}_{;\alpha}=0$.
However, at fist sight the RHS of the modified field equation in
(\ref{eqn:G2}) would appear not to be divergence-free in that the
expression that replaces the gravitational constant $G$ seems to
be time dependent, i.e.

\begin{equation}\label{eqn:divergence}
  \frac{\partial }{{\partial x^0 }}\left(
{\frac{3 c^2}{{8 \pi R^2 \bar \rho }}} \right) \ne 0
\end{equation}

It would appear that either we have to forego the divergence-free
nature of the original Einstein equation, or we have to `adjust'
the new equation in some way in order to retain this desirable
property. The latter approach was used in the Brans-Dicke
scalar-tensor theory \cite{dicke:mach}, and subsequently in the
Canuto scale-covariant theory \cite{canuto:1977}. Although these
fixes solved the immediate problem by restoring the
zero-divergence property, this was done at the expense of the
overall elegance of the solution, and ultimately its ability to
make any useful predictions.  There is, however, another more
radical solution to this problem.

In order to preserve the principle of general covariance, and
still retain the features of the gravitational field equations, we
must conclude that the period of an atomic clock is decreasing in
proportion to the cosmological scale factor. Accordingly, we
substitute for $R\equiv ct$ in (\ref{eqn:divergence}) with
$cn\tau$, where $n$ is the time in atomic time units, and $\tau$
is the period of our atomic clock. Since $n\propto a = 1/\tau$, we
see that the expression in (\ref{eqn:divergence}) is invariant
with respect to $x^0\equiv \tau$, and hence possesses zero
divergence.

\par This leads to a somewhat bizarre picture of time and
space. From the perspective of an observer in the cosmological
reference frame the Universe would appear to be static and closed.
Any concentrations of matter in the Universe would appear to be
shrinking in size. For an observer, such as ourselves, linked to
the atomic reference frame, the Universe will appear to be flat
and expanding, with the horizon receding at the speed of light.
Our perception of time as a continuum is an illusion, and the time
co-ordinate that we are used to is perhaps better described as
subjective time. (This notion is not entirely novel; something
similar was proposed by Jeans in 1931 \cite{jeans:1931}). The
concept of two kinds of time - cosmological time and atomic time -
is similar in some respects to the dual timescales postulated by
Milne in his kinematic theory of gravity \cite{milne:1938}.

\par It is interesting to look at the conclusions that would be
drawn by an experimenter who was somehow able to measure lengths
using the cosmological scale appropriate to General Relativity,
whilst still only having access to clocks based on atomic time.
Being unaware that their clock was speeding up relative to
cosmological time, such an observer attempting to measure the
speed of light with sufficiently accurate apparatus would conclude
that $c$ was decreasing with time. This is analogous to the
concept of varying speed of light cosmology (see
\cite{magueijo:1998} for example).

\section{Consequences and Predictions}\label{sec:consequences}
The definition of a Universe provided in Section
\ref{sec:scalefree}, together with its associated gravitational
field equations and resultant cosmic dynamics, describes what is
essentially a new paradigm.  Arguably, if this can be shown to be
valid, then it would supplant the current standard model that
includes the Big Bang theory and provides part of the rational for
pursuing a theory of Quantum Gravity.  This section highlights
some of the principal consequences that would follow from this new
paradigm.

\subsection{The Age of the Universe}
The fact that this model describes a universe with a linearly
increasing scale factor, such that $H=1/t$, means that there is
now no disparity between the measured value of the Hubble constant
and the age of the universe as derived from astrophysical and
geophysical data, i.e. about $15 \times 10^9$ years.

\subsection{Solving the Big Bang problems}

\subsubsection{The flatness problem}
The dynamical equations in Section \ref{sec:dynamics2} clearly
show that, for the Universe as a whole, the mean energy density is
maintained at the critical value by a form of negative feedback
mechanism. Under such circumstances, any small deviation of
$\Omega$ from unity would result in an apparent acceleration or
deceleration of the expansion rate so as to bring the system back
to its equilibrium state.

\subsubsection{The horizon problem}
The QSU model requires the Universe to be spatially closed, with
the topology of a 3-sphere. Because the expansion rate is
constant,the horizon distance will always be equal to the radius
of the 3-sphere that defines the observable Universe. As a result
of this coincidence of horizon distance and the radius of the
observable Universe, all regions within the Universe will have
been in causal contact with each other at time $t=0$. This
accounts for the observed homogeneity of the Universe, and
provides an elegant solution to the horizon problem.

The smoothness problem, i.e. accounting for the perturbations in
matter density that give rise to structure formation, is not
explained directly by the QSU model. The variations in matter
density that are required as a prerequisite for galaxy structure
formation must therefore arise from some other mechanism.

\subsubsection{The cosmological constant problem} Since the QSU
model accounts for the observed dynamics of the Universe without
the requirement for a non-zero $\Lambda$, this constant can
validly be omitted from the gravitational field equations. Hence
the problem of how to explain a small, but non-zero, $\Lambda$
disappears. Similarly, the linearity problem rased in Section
\ref{sec:bigbang} also ceases to be relevant as the QSU model
results in a linearly expanding universe without the need to
achieve a delicate balance between $\Omega_\Lambda$ and
$\Omega_M$. As was pointed out earlier, the scale factor of a
universe with $\Omega_M \simeq 0.28, \Omega_\Lambda \simeq 0.72$
increases in a very similar way to the linear expansion predicted
in Section \ref{sec:dynamics2}.  The main divergence between the
two models occurs at redshifts in the range $0.8<z<1.8$.  In
\cite{booth:2001} it was shown that results from the only
supernova yet discovered in this redshift range tend to favour the
QSU model over the standard model.

\subsection{Variations in the gravitational
constant}\label{sec:varyingG}
In Section \ref{sec:Newton} it was
shown that Newton's gravitational constant, $G$, would be
proportional to $t$ in the reference frame of an observer on the
atomic scale. The most accurate methods currently used to measure
\( {\dot G/G} \) are generally based on the principle of measuring
changes in planetary orbits within the solar system using radar
ranging techniques. However, any calculation of \( {\dot G/G} \)
based on distance measurements that are ultimately derived from
atomic scale phenomena will generate a null result.  This is due
to the fact that time measured by any atomic or gravitational
clock will be changing at the same rate as the distance to be
measured. Suppose at time $t_0$ and scale factor $a_0$ the
measured distance is $2r_0=cn\tau_0$, where $\tau_0$ is the period
of an atomic clock and $n$ is the number of clock ticks between
the emission of a radar signal and the reception of its
reflection. At some future time when the scale factor has
increased to $a$, the measurement is repeated.  If $G(t) \propto
a$ then the distance to the planet will have decreased so that
$r=r_0a_0/a$.  However, the period of the atomic clock will also
have decreased by the same proportion, with $\tau=\tau_0a_0/a$.
Consequently, the measured elapsed time for the radar signal round
trip will still be $n$ ticks, i.e. there will be no apparent
change in distance and therefore no change in the calculated
gravitational constant.

\par In order to verify that $G$ does vary over cosmological timescales
it is necessary either to measure its value directly using a
Cavendish type experiment, or to turn to evidence from geophysical
and astrophysical measurements. Since Cavendish experiments can
currently only achieve accuracies of one part in $10^{-6}$, these
are not capable of detecting changes in $G$, which will be of the
order of the Hubble factor, i.e. one part in $10^{-11}$ per year.
We must therefore look to the other sources for indirect evidence
of a time-varying $G$.  Fortunately, a useful source of
geophysical data does exist in the form of the Earth's fossil
record, which can be used to track changes in the length of the
Earth's day over geological timescales.  Analysis of this data in
\cite{arbab:2001} shows that the rotation of the Earth has slowed
down since the planet was formed, due to tidal interaction with
the Moon.  Conservation of angular momentum of the Earth-Moon
system dictates that there must be a corresponding increase in the
orbital angular momentum of the Moon.  Conventionally, this would
be achieved by an increase in the radius of the Moon's orbit, and
a corresponding decrease in the Moon's angular velocity, together
with a lengthening of the sidereal month.  However, observational
data suggests that the Moon is in fact accelerating in its orbit,
such that the change in the length of the lunar month is
consistent with a time varying gravitational constant, with $G(t)
\propto t$.

\subsection{Primordial nucleosynthesis}\label{sec:nucleosynthesis}
Arguably, one of the few successes of the standard Hot Big Bang
(HBB) model is its ability to predict the abundances of the light
elements resulting from primordial nucleosynthesis. If the QSU
model is to be of any use then it must also give predictions that
are consistent with observational data. The QSU model implies that
the observed power of the CMB is due to a relatively small number
density of energetic photons rather than a very large number
desnsity of low energy photons. If we assume that the CMB photons
originally had energies of $\sim 1 MeV$, corresponding to a
temperature of $\sim 10^{10 \circ} K$, and that these photons have
been redshifted to the currently measured temperature of $2.7^
\circ K$, this implies an energy loss of the order of $10^{10}$
according to the standard theory. Since the currently observed CMB
photon number density, calculated according to the standard
theory, happens to correspond to $\eta_B\equiv n_B/n_\gamma \sim
10^{-10}$ baryons per photon, it follows that $\eta_B \simeq 1$ in
the QSU model. This is an encouraging result in that it resolves
the entropy problem: there is no a priori reason why the entropy
of the universe should be as high as the value determined by the
HBB model.  However, it presents a potential difficulty in
reconciling observed light element abundances with an $\eta_B$ of
the order of unity.

\par According to the HBB model, the deuterium abundance in the
Universe is particularly sensitive to the baryon-photon ratio,
$\eta_B = 2.7 \times 10^{-8} \Omega_B h^2$.  The HBB model
requires a value of  $\eta_B$ corresponding to $\Omega_B h^2 \sim
0.02$ to give the observed Deuterium abundance of about $10^{-4}$.
However, it has been pointed out by Aguirre in \cite{aguerre:2001}
that the set of apparently arbitrary  parameters that defines the
standard HBB model, including $\eta_B, \Lambda$, etc, constitutes
merely one point in parameter space that can lead to a universe
compatible with human life.  It is, in principle, possible that
other combinations of the same parameters will also give rise to
viable universes. As an example, the case of the classical Cold
Big Bang (CBB) is examined.  This model is characterised by a
baryon-photon ratio of $\eta_B \simeq 1$. Detailed numerical
simulations of primordial nucleosynthesis reactions
\cite{aguerre:1999} show that, with the appropriate values for the
lepton-baryon ratio, $\eta_\lambda$, the CBB model is indeed able
to create levels of metallicity that are compatible with those
currently observed in the intergalactic medium.

\par The current CMB photon energy density is approximately
$10^{-3}$ of the observed energy density due to baryonic matter.
Taking these two observations together, this suggests that we are
looking for a nucleosynthesis model that results in one photon per
baryon, with an energy approximately $10^{-3}$ of the proton rest
mass energy.  The most obvious scenario is that of neutron decay,
first proposed by Gamow \cite{gamow:1}.  Initial studies of the
evolution of a cold neutron Universe have been carried out using
numerical simulation models.  These show that the initially cold,
dense, neutron cloud heats up by means of $\beta$ decay to form a
hot proton-neutron-electron plasma at a temperature of $\sim
10^{11  \circ} K$.  At $\sim 10^{9  \circ} K$ a range of fusion
reactions become energetically favorable, and lead to the
formation of deuterium, tritium, helium and other light elements,
as in the standard Big Bang nucleosynthesis models
\cite{wagoner:1967}.  The main differences between the  QSU model
and the standard Big Bang is that in the former, photons play a
negligible role in the exchange of energy between particles.  The
fact that the expansion rate is much slower also has a significant
impact in that there is more time for fusion reactions to take
place, and therefore an increased probability of synthesizing
heavier elements than would be the case with the standard model.
Nucleosynthesis in a linearly expanding Universe has also been
studied by Lohiya in \cite{lohiya:1998}.

\par One of the most
important consequences of nucleosynthesis in the  QSU model is
that the primordial baryonic matter is not able to cool down as
the Universe expands, since there are insufficient photons to
remove the entropy generated by the neutron decay and nuclear
fusion processes.  The primordial hydrogen and helium molecules
will therefore remain in an ionized state indefinitely.  This may
explain why intergalactic gas clouds are currently observed to be
ionized.  Another feature of this model is that the $\beta$ decay
process that causes the primordial universe to heat up will give
rise to scale-invariant differences in temperature, due to the
statistical nature of the neutron decay reaction. Starting from a
perfectly isotropic and homogeneous state, this mechanism is
therefore able to account for the observed scale-invariant
temperature fluctuations in the CMB, which are explained by
inflation in the standard Big Bang model.

\subsection{The Large Number Hypothesis} A dimensionless quantity
known as gravitational structure constant can be defined as the
ratio of the electrostatic forces between two adjacent charged
particles, e.g. protons, to the gravitational force between the
particles.

\begin{equation}\label{alphaG}
\alpha _G  =  \frac{{Gm_p ^2 }}{{\hbar c}}{\rm } \approx {\rm
5}{\rm .9} \times {\rm 10}^{{\rm  - 39}} 
\end{equation}

Standard cosmological theories provide no obvious explanation for
such a vast disparity between the forces of gravity and
electromagnetism.  In 1938 Dirac  \cite{dirac:1938} noted that the
dimensionless quantity \( {1 \mathord{\left/
 {\vphantom {1 {\alpha _G }}} \right.
 \kern-\nulldelimiterspace} {\alpha _G }}
\) was approximately equal to the present age of the Universe
measured in atomic time units (where 1 atomic time unit = \(
{\hbar  \mathord{\left/
 {\vphantom {\hbar  {m_p c^2  \approx }}} \right.
 \kern-\nulldelimiterspace} {m_p c^2  \approx }}10^{ - 24}
\) secs).  If this relationship were to be valid for all epochs
then this implies that \( {1 \mathord{\left/
 {\vphantom {1 {\alpha _G }}} \right.
 \kern-\nulldelimiterspace} {\alpha _G }}
\) must be proportional to the age of the Universe, and therefore
that \( G\left( t \right) \propto {1 \mathord{\left/
 {\vphantom {1 t}} \right.
 \kern-\nulldelimiterspace} t}
\).  This postulate formed the basis of Dirac's Large Number
Hypothesis (LNH), which has subsequently provided the inspiration
for a number of alternative cosmological theories. (It is worth
noting that this formulation of the LNH is equivalent to the
expression \( {{G\rho _0 } \mathord{\left/
 {\vphantom {{G\rho _0 } {H_0^2  \simeq 1}}} \right.
 \kern-\nulldelimiterspace} {H_0^2  \simeq 1}}
\) of Mach's principle).

\par In looking at some of the implications of this relationship for
observers in the atomic reference frame it is helpful to express
the mean gravitational energy density of the Universe in terms of
the baryon number $N$, and the mean baryon mass, which we shall
take to be the proton mass $m_p$.  (Note that this implies, but
does not require, that any missing mass in the Universe is
baryonic in nature rather than in the form of other more exotic
entities).
\begin{equation}
    \bar \rho  = \frac{{3Nm_p }}{{4\pi R^3 }}
\end{equation}
The expression for $G$ in (\ref{eqn:G}) can therefore be written
as
\begin{equation}\label{Gsub3}
G = \frac{{R(t)c^2 }}{{6Nm_p }}
\end{equation}

If we now substitute for $G$ in equation (\ref{alphaG}) for the
gravitational structure constant $\alpha_G$, we find that

\begin{equation}
    \alpha _G  = \frac{{R(t)cm_p }}{{6N\hbar }} \label{eqn:alphag1}
\end{equation}
where $R(t)$ is the apparent radius of the Universe in the atomic
reference frame, at subjective time $t$.  From this it can be seen
that $\alpha_G \propto t$ in our reference frame, i.e. the
strength of the gravitational interaction between particles will
increase over time in relation to their mutual electromagnetic
forces.  Combining (\ref{eqn:alphag1}) with the expression for
atomic time we find that

\begin{equation}
\alpha _G  = \frac{n}{N}
\end{equation}

where $n$ is the time in atomic time units. Although this very
simple result may at first seem somewhat surprising, it is perhaps
to be expected, since the baryon number $N$ is one of the few
dimensionless quantities to occur naturally in cosmology. (The
fact that $1/\alpha_G \approx n$ today is purely a coincidence).

\subsection{Planck Units}
We shall now examine the effects of recasting the expressions for
the Planck units using the formula for $G$ given in (\ref{Gsub3}),
and the de Broglie wavelength of a proton given by \( R \approx
\lambda _p  \approx {\hbar  \mathord{\left/
 {\vphantom {\hbar  {m_p c}}} \right.
 \kern-\nulldelimiterspace} {m_p c}}
\).

\paragraph{Planck Length}
Clearly with $G(t) \propto t$ the quantity known as the Planck
Length in (\ref{Lplanck}) will itself be a function of time such
that \( L_P (t) \propto \sqrt t \).  Substituting for $G$ using
(\ref{Gsub3}), and the expression for the atomic time unit, gives

\begin{equation}
  L_P = \lambda_p \sqrt{\frac{n}{N}}
\end{equation}

where $n$ is the time expressed in atomic time units, and $N$ is
the baryon number of the Universe. It is interesting to note that
at a time $n=N$ the Planck Length will have grown to a size such
that $L_P= \lambda_p$.  (Or conversely, in the cosmological frame,
the proton wavelength will have shrunk below the Planck Length).

\paragraph{Planck Time}
A similar set of expressions can be derived for the quantity known
as Planck Time in (\ref{Tplanck}), to give
\begin{equation}
T_P  = \frac{\lambda_p}{c} \sqrt{\frac{n}{N}}
\end{equation}

\paragraph{Planck Mass}
From Equation (\ref{Mplanck}) it is evident that the Planck Mass
\( M_p (t) \propto t^{ - {\textstyle{1 \over 2}}} \). Again,
substituting for $G$ using (\ref{Gsub3}), with \( R = \lambda _p
\) we find
\begin{equation}
M_P  = m_p \sqrt {\frac{N}{n}}
\end{equation}

And when $n=N$ we see that \( M_P  = m_p \).

\par Based on this analysis, the conclusion we must reach is that Planck Units
do not represent a fundamental measurement scale that becomes
relevant during the birth of the Universe and governs the realm of
Quantum Gravity. Rather, they are scale factor dependent
quantities which may shed some light on the behaviour of the
Universe in its dying moments.

\subsection{The ultimate fate of the Universe}
In the preceding section we saw that at a time $n=N$ (where $n$ is
the time in atomic time units, and $N$ is the baryon number of the
Universe), the evolution of the scale invariant Universe reaches a
state at which the Planck length is equal to the Compton
wavelength of the proton, and the Planck mass is equal to the
proton mass. Recalling that the Schwartzchild radius of a black
hole is given by
\begin{equation}
    R_S  = \frac{{2GM}}{{c^2 }}
\end{equation}

and substituting for $G$ using (\ref{Gsub3}), with $R(t)=N
\lambda_p$ and $M=m_p$, we find that
\begin{equation}
     R_S  = \lambda_p
\end{equation}

In other words, the scale factor of the Universe has evolved to
the point where the radius of the proton exceeds the Schwartzchild
radius corresponding to the proton mass.  (The term proton here is
used loosely to refer to whatever state baryonic matter may exist
in the extreme gravitational conditions prevailing at this epoch.
In practice it is more likely that protons and electrons will have
recombined into atomic hydrogen by this stage, which in turn may
have collapsed into neutrons in a reversal of the process
described in \ref{sec:nucleosynthesis} above). At this point the
Universe effectively comes to an end as all protons simultaneously
collapse into micro Black Holes - possibly to give birth to many
more baby Universes according to Smolin in \cite{smolin:1994}.

\section{Conclusions and discussion}\label{sec:conclusions}
We have seen that, by challenging the assumption that spacetime
curvature on a cosmological scale should be linked to the
behaviour of fields on an atomic scale, it is possible to
construct a gravitational field equation that is truly scale
invariant.  It has been shown that the cosmic dynamics resulting
from this formulation of gravity describe a quasi-static universe,
possessing a number of interesting properties.  It is argued that
the predicted behaviour of this QSU model provides a much more
elegant explanation for a number of astronomical observations than
does the standard HBB plus inflation model. Furthermore, the scale
invariance inherent in the QSU model renders Newton's
gravitational constant redundant, and with it the concept of a
uniquely defined Planck scale. This feature in itself has a number
of profound consequences across many areas of physics, some of
which will be discussed here.

The elimination of the initial Big Bang singularity, taken
together with the observation that the gravitational structure
constant must be changing over time in the QSU model, leads to a
very different picture of the early Universe.  For example, at the
epoch when matter fields initially come into existence, the
electromagnetic self-energy of a particle will be exactly equal to
the sum of the gravitational energy of the particle with respect
to every other particle in the Universe.  This should provide a
useful clue about the nature of the mechanism that determines
particle masses.

The concept of a uniquely defined Planck scale is one of the two
principal motivations for the pursuit of a quantum theory of
gravity, the other being the need for the curvature terms in the
gravitational field equation to be quantized in order to be
equivalent to the quantized matter fields. If the Planck factor is
removed, we need to ask whether there is still a need for a theory
of Quantum Gravity, at least in the form currently being sought.
There is no fundamental requirement that the gravitational field
should have an inherent quantum structure, and it may well be more
reasonable to think of the gravitational field as being quantised
as a consequence of the matter fields with which it interacts.

The QSU model also has a number of implications for the role of
time.  The static nature of the Universe, when viewed from a
cosmological reference frame, strongly suggests that at its most
fundamental level the Universe does not possess any dynamical
structure.  If this were to be the case then we would need to
reject the concept of time in canonical quantum gravity as being
the separation between successive foliations of a three
dimensional manifold.  The fourth dimension in General Relativity
would then take on a hyperspatial role, providing the additional
dimension in Euclidean space in which the three dimensional
manifold of the observable Universe can be curved.  Time in
Quantum Mechanics would continue to be a feature of the Lorentzian
metric that overlays Euclidean hyperspace.

Finally, one can speculate that by more clearly defining the
distinction between the realms of General Relativity and Quantum
Mechanics, we can actually move closer towards constructing a
paradigm that unifies the two theories.

\section*{References}
\bibliographystyle{unsrt}
\bibliography{Planck}

\begin{thebibliography}{10}

\bibitem{dicke:mach}
C.Brans and R.Dicke.
\newblock Mach's principle and a relativistic theory of gravity.
\newblock {\em Physical Review}, 124:925--935, 1961.

\bibitem{pervushin:2001}
V.Pervushin and D.Proskurin.
\newblock conformal general relativity.
\newblock {\em Preprint}, gr-qc/0106006.

\bibitem{guendelman:2001}
E.I.Guendelman.
\newblock scale invariance, mass and cosmology.
\newblock {\em Preprint}, gr-qc/9901067.

\bibitem{mannheim:1989}
P.Mannheim and D.Kazanas.
\newblock conformal cosmology.
\newblock {\em Astrophysical J.}, 342:635, 1989.

\bibitem{mannheim:1990}
P.Mannheim.
\newblock conformal cosmology with no cosmological constant.
\newblock {\em Gen. Relativ. Gravit.}, 22:289, 1990.

\bibitem{hoyle}
G.Burbage F.Hoyle and J.Narlikar.
\newblock {\em A Different Approach to Cosmology}.
\newblock Cambridge University Press, 2000.

\bibitem{canuto:1977}
V.Canuto, P.J.Adams, S.H.Hsieh, and E.Tsiang.
\newblock Scale-covariant theory of gravitation and astrophysical applications.
\newblock {\em Physical Review D}, 16(6):1643, 1977.

\bibitem{canuto:1978}
V.Canuto, H.S.Hsieh, and P.J.Adams.
\newblock Mach's principle, the cosmological constant, and the scale-covariant
  theory of gravity.
\newblock {\em Physical Review D}, 18(10):3577, 1978.

\bibitem{goobar}
L.Bergstrom and A.Goobar.
\newblock {\em Cosmology and Particle Astrophysics}.
\newblock Wiley, 1999.

\bibitem{hamuy:1996}
M.Hamuy et~al.
\newblock {the absolute luminosities of the Calan/Tololo Type Ia supernovae}.
\newblock {\em Astrophysical Journal}, 112:2391, 1996.

\bibitem{jm:1}
J.Magueijo and K.Baskerville.
\newblock Big bang riddles and their revelations.
\newblock {\em Phil. Trans. R. Soc. A}, 357(1763):3221, 1999.

\bibitem{reiss:supernovae}
A.G.Reiss.
\newblock observational evidence from supernovae for an accelerating universe
  and cosmological constant.
\newblock {\em AJ}, 116:1009, 1998.

\bibitem{hawking:1970}
S.W.Hawking and R.Penrose.
\newblock The singluarities of gravitational collapse and cosmology.
\newblock {\em Proc. Roy. Soc. London}, A314:529, 1970.

\bibitem{barbour:mach}
J.Barbour and H.Pfister.
\newblock {\em From Newton's Bucket to Quantum Gravity}.
\newblock Birkhäuser, 1995.

\bibitem{callender}
C.Callender and N.Huggett.
\newblock {\em {Physics Meets Philosophy at the Planck Scale}}.
\newblock Cambridge University Press, 2001.

\bibitem{butterfield}
J.Butterfield.
\newblock {\em {The Arguments of Time}}.
\newblock Oxford University Press, 1999.

\bibitem{booth:2001}
R.A.J.Booth.
\newblock Machian general relativity: a possible solution to the dark energy
  problem and an alternative to big bang cosmology.
\newblock {\em gr-qc/0106007}, 2001.

\bibitem{barrow:1999}
J.D.Barrow.
\newblock Varying g and other constants.
\newblock {\em gr-qc/9711084}, 1999.

\bibitem{jeans:1931}
J.Jeans.
\newblock Evolution of the universe.
\newblock {\em Nature}, 128:703, 1931.

\bibitem{magueijo:1998}
A.Albrecht and J.Magueijo.
\newblock a time varying speed of light as a solution to cosmological problems.
\newblock {\em astro-ph}, 9811018.

\bibitem{milne:1938}
E.A.Milne.
\newblock Kinematic relativity.
\newblock {\em Proc. R. Soc. A}, 165:351, 1938.

\bibitem{arbab:2001}
A.I.Arbab.
\newblock {Determination of the cosmological parameters from the Earth-Moon
  system evolution}.
\newblock {\em Preprint}, astro-ph/0107024.

\bibitem{aguerre:2001}
A.Aguirre.
\newblock the cold big bang cosmology as a counter-example to several anthropic
  arguments.
\newblock {\em Preprint}, astro-ph/0106143.

\bibitem{aguerre:1999}
A.Aguirre.
\newblock cold big bang nucleogenesis.
\newblock {\em ApJ.}, 521:17--29, 1999.

\bibitem{gamow:1}
G.Gamow.
\newblock Evolution of the universe.
\newblock {\em Nature}, 162:680--2, Oct 1948.

\bibitem{wagoner:1967}
R.V.Wagoner, W.A.Fowler, and F.Hoyle.
\newblock Primordial nucleosynthesis.
\newblock {\em Ap.J.}, 148:3, 1967.

\bibitem{lohiya:1998}
D.Lohiya et~al.
\newblock Nucleosynthesis in a simmering universe.
\newblock {\em Preprint}, gr-qc/9808031.

\bibitem{dirac:1938}
P.A.M.Dirac.
\newblock The large number hypothesis.
\newblock {\em Proc.R. Soc. A}, 165:199, 1938.

\bibitem{smolin:1994}
L.Smolin.
\newblock {The fate of Black Hole singlularities and the parameters of the
  standard models of particle physics and cosmology}.
\newblock {\em gr-qc/9404011}, 1994.

\end{thebibliography}

\end{document}